**The evolution of asymmetrical regulation of physiology is central to aging**

Mirre J P Simons[1] & Marc Tatar[2]

[1] *School of Biosciences, University of Sheffield, Sheffield, UK*
[2] *Department of Ecology, Evolution and Organismal Biology & The Center for the Biology of Aging, Brown University, Providence, RI, USA*

**Abstract**

The evolutionary biology of aging is fundamental to understanding the mechanisms of aging and how to develop anti-aging treatments. Thus far most evolutionary theory concerns the genetics of aging with limited physiological integration. Here we present an intuitive evolutionary framework built on how physiology is regulated and how this regulation itself is then predicted to age. Life has evolved to secure reproduction and avoid system failure in early life, and it is the physiological regulation that evolves in response to those early life selection pressures that leads to the emergence of aging. Importantly, asymmetrical regulation of physiology will evolve as the Darwinian fitness costs of loss of regulation will not be symmetrical. When asymmetrical regulatory systems break during aging, they cause physiological function to drift towards the physiological range where costs of dysregulation are lowest, rendering aging directional. Our model explains many puzzling aspects of the biology of aging. These include why aging appears (but is not) programmed, why aging is gradual yet heterogeneous, why cellular and hormonal signaling are closely related to aging, the compensation law of mortality, why trade-offs between reproduction and aging remain elusive, why longer-lived organisms show more signs of aging during their natural lifespans, and why longer-lived organisms can be less responsive to treatments of aging that work well in short-lived organisms. We provide predictions of our theory that are empirically testable. By incorporating physiological regulation into evolutionary models of aging, we provide a novel perspective to guide empirical research in this still growing field.

**A physiological perspective on the evolutionary biology of aging**

Evolution optimizes physiology across different levels of organization. Optimality theory has been influential in how we understand aging (Kirkwood, 1977; Kirkwood & Austad, 2000; Partridge & Barton, 1993), with trade-offs assumed to arise between the operation of physiological function in early life (i.e. reproduction, Höglund et al., 1998) with consequences in later life. Darwinian fitness is optimized at the largest difference between the benefits accrued at young ages relative to discounted costs at later ages – mortality and impaired fertility. However, physiological responses are also optimized by evolution within any age class. For example, optimality models explain why immune defenses at one age can be risky or somewhat self-destructive because the potential fitness costs to not mount an immune response can be catastrophic (Nesse, 2005). We recently expanded these optimality models in the context of Alzheimer's disease to explain how activation and negative feedback on Aβ production may account for its accumulation with age (Tatar, 2022). Here, we expand this model and generalise it to the biology of aging (Figure 1). The key difference with our insight compared to other theories in aging is that we view early life trade-offs as the currency that is optimized by evolution, and argue the subsequent breakdown of the systems that regulate these trade-offs causes aging to emerge.

The fundamental basis of the evolution of aging is that selection is strongest in early life and decreases with age (Fisher, 1930). There is strong selection in early life to maintain, secure and optimize current physiological operation to survive and sustain reproduction. Indeed, in many if not all species (Bonnet et al., 2022), including our own (Nettle & Pollet, 2008), Darwinian fitness is highly zero-inflated. This means that many individuals in a population do not have any fitness. The physiological traits optimized by selection therefore do not only include traits that contribute to variation in reproductive output, but also include 'basic' physiology that sustains life to secure any reproductive output. Catastrophic failure is strongly selected against, as premature mortality or the early loss of fecundity incurs a loss of all fitness. Indeed, in a broader sense, alleles that have the highest geometric rather than arithmetic mean fitness are favored by natural selection (Orr, 2009). Thus, when selection optimizes physiology in young adults it does so by maximizing the *current* benefits compared to the *current* costs (Figure 1). Regulation of physiology is thus expected to be strongly selected to sustain and secure both reproduction and survival in *early* life. We argue here that it is this evolution of physiological regulation optimized to *early* life fitness outputs that is central to the emergence and properties of aging later in life.

In a basic optimality model (Figure 1) between the cost of not inducing a response ('cost of inertia') and the cost of inducing a response ('cost of induction') the optimal physiological response is where the sum of both costs are minimized (Nesse, 2005; Tatar, 2022). In the context of Alzheimer's disease, Tatar (2022) argued that such optimal physiological states are maintained by induction and feedback. In expanding this concept to aging in general, we ask here how such physiological regulation evolves and how regulatory systems subsequently age. We reason that early life selection pressure causes regulation to evolve with an asymmetry between induction and feedback, and this asymmetry determines how aging emerges. We show that this novel perspective explains many aspects of the biology of aging. We provide multiple novel testable predictions of our theory and suggest a focus on how young animals regulate physiology will reveal how we can negate the loss of regulation during aging.

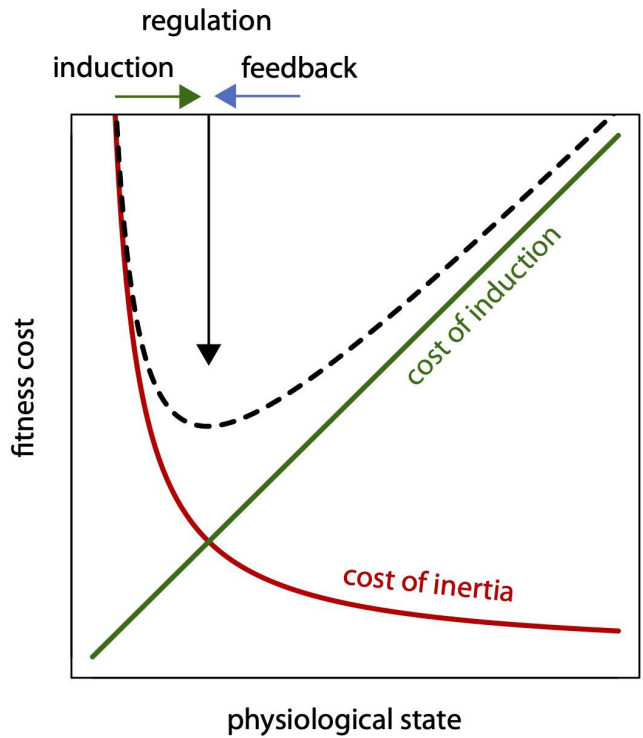

**Figure 1.** Physiological state is optimized by evolution based on cost functions of inducing the physiology (green line) and doing nothing (red line). The minimum of the sum of these costs (dashed black line) determines the optimum (arrow). On a physiological level regulation evolves with a combination of feedback and induction to retain the most optimal physiological state for fitness (green and blue arrows). Our theory concerns the breakdown of this regulation during aging. A loss of induction would lead to increased costs of inertia, a loss of feedback to increased costs of induction. The shape of the cost curves (green and red) determines the optimal response and the consequences of aging when regulation deteriorates over time. We argue that the resulting cost curves (dashed) as selected in early life are asymmetrical, and that this property makes it central to the later biology of aging. These cost curves can represent any aspect of physiology. Intuitively, and previously, interpreted in light of immunity: the cost of inertia (red) representing the cost of virulence and the cost of induction (green) representing the costs of mounting an immune response. Figure redrawn from Nesse, 2005 and Tatar, 2022.

**The basis of the evolutionary biology of aging**
Selection favors individual genotypes with the highest production, which crudely is the sum of their progeny in a lifetime. Intrinsic mortality, i.e. aging, together with extrinsic causes of mortality (such as accidents, disease) determine lifespan, which in turn with reproductive rate totals to lifetime reproductive success. Aging and reproduction are thus tightly linked, and it is no surprise that evolutionary trade-off models have been prevalent in the biology of aging (Cohen et al., 2020; Johnson et al., 2019; Kirkwood & Austad, 2000; Stearns, 1989). Trade-offs dictate that when resources are invested in one aspect of physiology, e.g. reproduction, they carry a cost as these resources cannot be used to support other aspects of physiology, e.g. healthy aging. However, support for such trade-offs is relatively limited, in naturalistic contexts (Winder, Simons, et al., 2025) and mutants exist that increase both reproduction and survival (Briga & Verhulst, 2015; Lind et al., 2021; Wit et al., 2013; Yamamoto et al., 2021). In Medawar's mutation accumulation (Medawar, 1952) and Williams' antagonistic pleiotropy (Williams, 1957) theory of aging, late acting deleterious mutations accumulate in the genome over evolutionary time because selection reduces with age and/or is counterbalanced by early life benefits. Although there is some evidence for both mutation accumulation (Cui et al., 2019; Long & Zhang, 2019; Rodríguez et al., 2017; Wong & Holman, 2023; Yampolsky et al., 2000) and antagonistic pleiotropy (Long & Zhang, 2023; Parker et al., 2020; Rodríguez et al., 2017), effects are small and distributed across many loci. The disposable soma theory (Kirkwood, 1977) represents a physiological embedding of antagonistic pleiotropy which could explain why aging would evolve as a highly polygenic trait. However, the physiological mechanisms that drive aging have yet to be integrated in any theory (Mc Auley, 2025).

**The evolution of the regulation of physiology is central to life**

Physiology, including molecular cell machinery, is generally regulated by induction and feedback, which generates a state of homeostasis (Billman, 2020; Giordano, 2013; Modell et al., 2015). The term homeostasis is misleading in that stasis itself is rare in life. Rather, regulation of physiology allows organisms to respond optimally to the environment (Rattan, 2008; Sterling, 2012; Yates, 1994). As life is not static, the internal state of an organism will fluctuate depending on the environment. The environment here includes stressors, for example infection (Tatar, 2022), but also opportunity, for example food availability (McCracken et al., 2020). Furthermore, internal regulatory mechanisms do not provide an instant return to a static state, they rather operate to maintain a secure or optimal operational range for the organism (Sterling, 2012). Regulation of physiology (Cohen et al., 2012a, 2012b, 2022; Secor & Diamond, 1998) is optimized by evolution in two ways. First, physiology is tightly regulated so that the organism can operate at the physiological state that produces most offspring during early life. Second, physiology evolves to avoid catastrophic failure securing both fecundity and survival in early life. An important consequence of this is that regulation of physiology is expected to be both strong and asymmetrical. Stronger and more layers of regulation are expected to evolve when aspects of physiology that are tightly linked to early life fitness are secured (Figure 2). In analogy to optimality models (Figure 1) and signal theory (Nesse, 2005), a skew in the cost curve on either end of the optimum is expected. Complete symmetry in physiological regulation is improbable in life. Asymmetry in the regulation of physiology that evolves to secure early life fitness determines how the system will subsequently age. When integrity is lost in regulatory systems with age this results in directional aging towards the physiological range with least regulation (Figure 2).

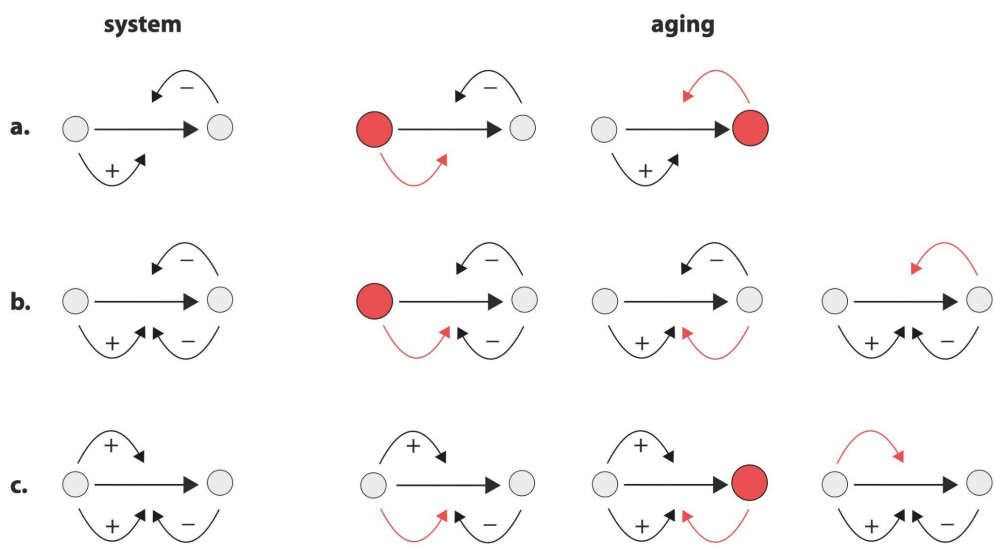

**Figure 2.** A simple regulatory system of moving (straight arrow) from one state to another (circles), with positive and negative feedback (curved arrows) regulating levels within a range that secures reproductive success. Biological systems (a) consist of layers of regulation that are asymmetrical (b, c). We argue that such asymmetrical regulation is inherent to life, and its selection in early life leads to the emergence of directional dysregulation during aging. In its simplest form, when individual components of regulation have equal chances of breaking with age (in red), systems age asymmetrically (b, c) dysregulating towards the side of least regulation (red fills).

**Why the regulation of physiology is central to the biology of aging**

Physiological regulation in the form of induction control, often coupled with positive and negative feedback loops, is a fundamental aspect of all life. There are many ways of modeling physiological regulation. One intuitive way is to imagine fluctuations in the internal state of the organism as a moving pendulum (Moldakozhayev & Gladyshev, 2023; Novák & Tyson, 2008; Vasileiou et al., 2019). Regulatory mechanisms push back on either end of the pendulum to maintain a certain operational range (Figure 3). Layers of regulation operate to then maintain the system within an optimal range also in the face of environmental perturbations. The pendulum model captures that work is required to achieve a physiological state. Gravity returns a pendulum to its resting state, but life operates outside of the space of least resistance and requires energy (or "work") to maintain a physiological state compatible with life. There is thus always a level of induction of the system towards a state at which life can flourish (Figure 3). This requirement of life renders the regulation of physiology inherently asymmetrical. Other considerations, from signal detection theory suggest similar asymmetrical cost curves (Figure 1). The consequences of threats to physiology often far outweigh the costs of defenses leading to an evolutionary optimum of seemingly overresponsive defenses, e.g. a high frequency of false alarm (Nesse, 2005). Regulatory systems of physiology are thus expected to both be strong and asymmetrical.

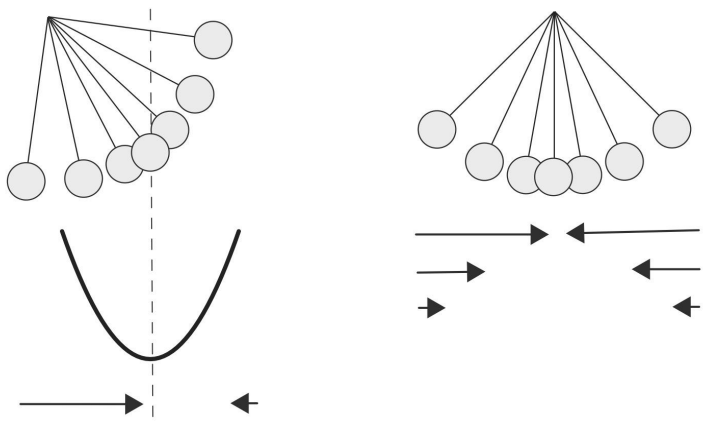

**Figure 3.** Regulation of physiology will be asymmetrical (left panel). The physiological state compatible with life is outside the resting state and therefore always requires work. The force required to move a system is dependent on the distance moved from its resting state. To acquire a physiological state that is compatible with life (dotted line) and subsequently is optimized for reproduction and survival assurance (bell curve) will therefore be asymmetrical (arrows). In addition, to achieve a mean physiological state physiological regulation evolves to maintain physiology within a certain range to correct for environmental perturbations in state (right panel). Tighter and more complex regulation is expected to evolve when a narrow range of physiological range results in highest Darwinian fitness, to counteract external and internal perturbation.

**Asymmetrical costs and the evolution of physiological regulation**

Aging is the attrition of biological systems, affecting every corner of biology, at differential rates and consequences to the organism. The evolution of aging is rather the opposite as aging itself is not the target of natural selection. Aging is the emergent phenotype resulting from the evolution of robustness against deterioration in the face of entropy and time. In this respect we can ask how we expect regulation of a physiological system to evolve and then subsequently age. We suggest that the appearance of late-age increments in functioning (i.e. aging) is not the delayed consequence of deleterious effects from the costs of early life benefits (as in strict antagonistic pleiotropy, Austad & Hoffman, 2018); they are a manifestation of how costs and benefits are managed through physiological regulation. During aging, physiological regulation systems degrade and organisms now more often venture into a physiological state that is damaging to the soma. As in a tug-of-war, equal forces at young ages keep the flag static. With age, loss of regulatory control permits inductive and feedback forces to move the flag, and only in those age-specific times do we recognize the marginal costs as "aging". Loss of physiological regulation during aging appears heterogeneous as it can result in three main outcomes, increased expression, decreased expression and variation in the expression of a trait (Figure 2, 4).

When the costs of dysregulation are asymmetrical, evolution is expected to favor selection to avoid the steepest costs. We expect to see additional redundancy of regulation operating to avoid the steeper side of the cost curve in the face of environmental perturbation and aging (Figure 2, 4). This redundancy can take many shapes, e.g. backup systems to sustain negative feedback and upregulation of positive feedback upon a loss of signal (self-enhancing). Crucially however, we expect biological systems to evolve to 'permit' consequences of aging to damage the feedback machinery on the side where most physiological costs occur should the organism leave its optimal operational space. Physiological regulation evolved for the fitness benefits in early life leads to aging that can be viewed as a double-edged sword of Damocles with one end sharper than the other.

We can even expect fail-safe mechanisms to evolve that actively force (e.g.) physiology in a certain direction upon environmental perturbation. Such mechanisms would avoid catastrophic death (or loss of fecundity) whilst sacrificing current functioning. When such regulatory systems break during aging, they can cause similar or even larger loss of function. Animals are not cars, single units not connected on an engineering level, e.g. brake failure is not connected to the gas pedal, allowing you to still accelerate even when your brakes have failed. Animals have evolved to be spaceships, with interlocked heavily redundant systems, if one system fails, the next system prevents catastrophe. A loss of physiological regulation with age leads to aging appearing as programmed physiology analogous to developmental programs (Kowald & Kirkwood, 2016), even though it is exactly the opposite.

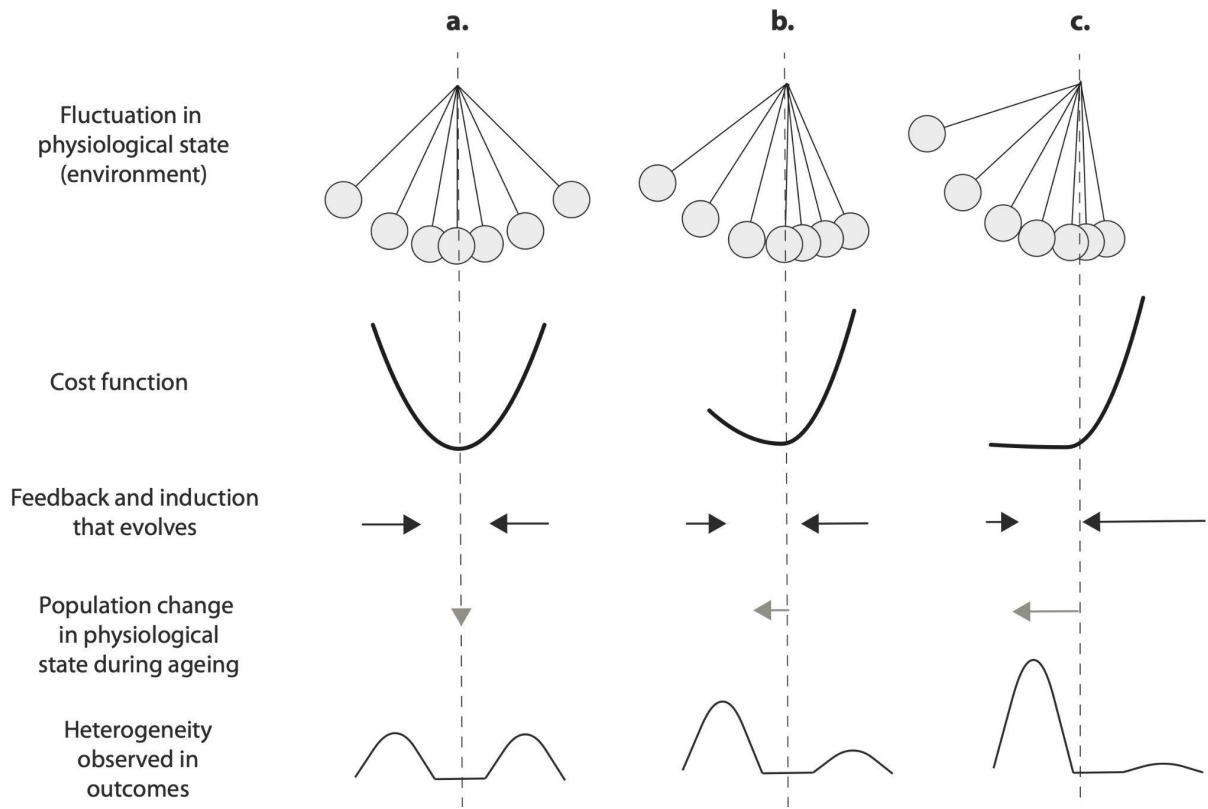

**Figure 4.** As life operates outside the resting state and environmental perturbations in state need to be buffered (Figure 3) physiological regulation evolves (arrows). This regulation will be asymmetrical as life requires work to allow the organism to operate at one side of the resting state (Figure 3). Furthermore, the consequences (solid line, cost function) of organisms operating outside their preferred physiological state (dashed line) are highly probable to be asymmetrical (a versus b & c). Note, as individuals age it is probable this cost function changes in shape and steepens when interconnected physiology fails. The cost functions here are Darwinian fitness costs, which are weighted more heavily to early life as this is where fitness is acquired or lost in the case of failure. The strength and redundancy of the feedback and induction mechanisms that evolve in response are asymmetrical, optimising the cost function (b, c). When the organism ages (Figure 2), the organism's state will drift towards the least regulated, least costly outcome, resulting in reduced functioning. This transition can be interpreted as frailty. When the cost function does not change substantially during aging, heterogeneity in outcomes, for example causes of death, will be increasingly biased to one direction. Note, however, a substantial proportion of the population is still predicted to succumb to the physiological state related to the steepest end of the cost curve, as a loss of regulation occurs during aging.

**Why aging appears programmed**

The reason why decay during aging is so canalized, is because physiological regulation is so strong. For example, cellular and hormonal signaling are tightly regulated by positive and negative feedback. This feedback is so strong that when organs producing key hormones, such as insulin, cortisol, estrogen and testosterone are partially removed, hormone levels recover (Burn et al., 2017; Cramer et al., 2023; Menge et al., 2012; Petersen et al., 1999). Cellular signaling pathways are also robust to perturbation (Billing et al., 2019; El-Brolosy & Stainier, 2017, 2017; Uda et al., 2013). When these highly regulated systems age, they fall apart in an apparently highly regulated fashion. A loss of one level of feedback is buffered by other interlocked regulatory systems, and such compensation slowly fails during aging. In addition, the least connected parts of these interconnected regulatory systems when damaged will cause distinct phenotypes as they are not effectively buffered, etching distinct aging phenotypes into how the organismal system fails over time.

**Why aging is so gradual**
Aging on an individual and population level is surprisingly gradual, with performance slowly reducing, and death preceded by a step-wise deterioration of all physiology. Our perspective on how central the regulation of physiology is in determining the biology of aging explains this observation. The whole organism when it ages is expected to slowly drift towards the side of physiological regulation where fitness costs are lowest during early life. Indeed on a whole organism level, including humans, this slow drift towards reduced function is observed and was recently termed 'mallostasis' (Pridham et al., 2024; Pridham & Rutenberg, 2023).

We predict aging will lead to reduced output of the physiological system if the fitness costs of over-induction are more costly during early life, as negative feedback starts to outcompete induction during aging. This scenario would fit with biology in which induction would lead to catastrophic damage, for example excessive proliferation during early life of the stem cell niche, with the cost of reduced stem cell activity during aging (Goodell & Rando, 2015). Or conversely, aging can lead to excessive activation when there is a strong selection pressure in early life to maintain activation, as then positive feedback starts to outcompete negative feedback when the system ages. Such a scenario fits with the hyperfunction theory of aging (Blagosklonny, 2021; Gems, 2022) and the developmental theory of aging (de Magalhães, 2012; de Magalhães & Church, 2005). The important distinction between those theories and our novel perspective is that we state that the dysregulation that emerges during aging is a direct result of how physiology is regulated during early adulthood. Aging of this highly regulated physiology is the emergent phenotype that appears to be regulated in itself, but this is a mere consequence of strong selection in early life.

> *We propose that the evolution of regulation of physiology is fundamental to aging and provides a physiological framework to the evolutionary biology of aging. This novel perspective explains several key observations in the field of aging, ranging from demography, to physiological and molecular mechanisms, and a new perspective on anti-aging treatments.*

**Cellular and hormonal signaling in aging**
We argue that it is no coincidence that hormonal and cellular signaling pathways have long since been associated with the biology of aging. Many of the first observations of mutants with prolonged longevity mapped to such signaling pathways. *Daf-2* was the first identification of a locus that could extend longevity, and its discovery was responsible for a paradigm shift by demonstrating the malleability of aging (Kenyon et al., 1993). Reductions in signaling (e.g. growth hormone / insulin signaling and mTor signaling) are associated with increased health and longevity (Garratt et al., 2016, 2017). A large effort is currently underway to understand the how, the inner workings of the cell and animal physiology associated with these pro-longevity pathways (Mannick & Lamming, 2023; Smulders & Deelen, 2024; Tatar et al., 2003). Why, in an evolutionary sense, a reduction in signaling strength is associated with an increase in longevity has been explained almost exclusively by a trade-off model, forcing investment away from reproduction towards survival (Bartke et al., 2013; Husak & Lailvaux, 2022; Lind et al., 2016; Schmeisser & Parker, 2019; Tatar et al., 2003).

Here we argue rather that the optimal signaling level itself balances current benefits with future intrinsic mortality. When reduced signaling has steeper immediate costs than enhanced signaling, regulation evolves asymmetrically with additional redundancy to push the system's state away from underactivation (Figure 3). When the system ages it is predicted to drift to an overactivated state. Indeed, such observations have been made for mTOR , and include humans (Markofski et al., 2015a). However, IGF-1 levels decrease with age in humans (Florini et al., 1985; Friedrich et al., 2010) and have a u-shaped relationship with all-cause mortality (Rahmani et al., 2022). Interestingly, however, IGF-1 levels show a skewed distribution with a long tail (in all studies included in the dose response in Rahmani et al. 2022). This means that on a linear dose-response curve, as in Rahmani et al. 2022, the population suffers most mortality due to excessively high IGF-1 levels. Thus even when signaling decreases during aging, dysregulation can still exhibit asymmetrical aging as we hypothesize. Unfortunately, limited investigation has gone into the distribution of signaling during aging and effects are also likely to be tissue specific (Baar et al., 2016). Anti-aging treatments that suppress signaling counteract the system's tendency to incur mortality costs at the enhanced signaling end. Still, at old age when signaling will be less well regulated, and thus more variable, there is a risk that the system will suffer from costs of a lack of induction. A loss of regulation during aging means that on both ends of the cost curve, even when the mean drifts, physiology wanders into the physiological ranges where costs are accrued. Such considerations could be especially important when anti-aging treatments are employed that suppress signaling in the very old.

**Biodemography of aging**
Simple but highly explanatory models (across different species and environments) of mortality rate are two-parameter models. One parameter describes the rate of exponential increase of mortality with age and the other an age-independent component (Kirkwood, 2015; Ricklefs, 2010a; Simons et al., 2013). Currently we lack an explanation of how these two demographic traits evolve and why. We also lack an explanation why these two demographic parameters evolve to be negatively correlated (compensation law of mortality) (Gavrilov & Gavrilova, 2001; Strehler & Mildvan, 1960), but see (Fedichev & Gruber, 2025). Our basic model of how the regulation of physiology leads to conceptually a similar dichotomisation through the asymmetrical fitness costs of physiological regulation. The steepest part of the cost landscape represents the age-independent risk of mortality. At most times the most steep costs are avoided as physiological regulation evolves that steers the organism away from the most costly physiological states. When damage accumulates with age, regulatory systems break down, and physiology drifts towards the lower end of the cost curve, leading to an exponential increase of mortality with age. The increased variability in physiological state that occurs with aging still causes deaths on both sides of the cost curve (Figure 4).

The age-independent mortality component in most demographic models of aging is a multiplier of the costs of mortality with age. Interestingly, this means that in absolute terms the amount of individuals that die from age-independent causes of mortality increases with age. This model fits with our model of physiological regulation that breaks due to aging. Aging will lead to increased variance in physiological state with age meaning that even though the system drifts towards the physiological state of least costs, deaths from age-independent mortality in absolute terms will still increase during aging. The relative force of both demographic patterns is determined by the relative risk posed by deviating out of the organism's operational space on either side. When one side of the physiological outcomes is avoided, i.e. when costs on this side are relatively steep, the other side is the physiological state that will disproportionately contribute to death and the state to which the organisms will drift to during aging. The asymmetrical fitness outcomes of physiological regulation make different outcomes of aging intrinsically tied and negatively correlated, exactly as in demography in the compensation law of mortality (Gavrilov & Gavrilova, 2001; Strehler & Mildvan, 1960).

**Trade-offs in reproductive effort**
The idea that reproduction is costly and directly impinges on aging, is largely founded in comparative biology: the observation that bigger, longer-lived animals tend to produce less young (Magalhaes et al., 2007; Ricklefs, 2010b). Within-species, evidence is limited, however. Although mutants that live longer often have reduced reproductive output, the link between reproduction and aging itself can be broken genetically and is not absolute (Flatt, 2011; Fletcher et al., 2015; Lind et al., 2021; Winder, Pick, et al., 2025; Yamamoto et al., 2021). Another area of research that revolves around costs of reproduction is parental care. When parental effort is manipulated directly within-species effects on subsequent mortality are small (Mitchell et al., 2024; Santos & Nakagawa, 2012; Schubert, 2009; Winder, Simons, et al., 2025). Animals are able to rear more offspring without clear costs on subsequent aging or survival. Effort in reproduction, as measured by parental care, is thus in natural situations underperforming. Energetic ceilings to parental effort have been proposed but these may only be rarely reached in the wild (Drent & Daan, 1980; Ricklefs et al., 1996; Simons et al., 2011; Speakman & Król, 2010). Our model predicts that animals have evolved to prioritize the avoidance of catastrophic failure of reproduction and of self, and thus explains why animals have evolved to underperform to avoid such failure. This interpretation fits with the observation that costs of reproduction only become apparent when animals are pushed outside of their natural physiological range (Winder, Simons, et al., 2025).

These insights are relevant as costs of reproduction have been central to explaining the evolution of the biology of aging (Kirkwood & Austad, 2000; Ricklefs, 2010b). However, the current evidence does not support this idea, and our model provides an alternative. Animals have evolved to avoid the steepest costs, and therefore reproduce at levels that almost always avoid catastrophic failure, i.e. death or loss of reproduction. Perhaps shorter-lived species should therefore show stronger costs of reproduction as they can afford to operate closer to physiological failure as they have fewer reproductive attempts. However, evidence relating a species' lifespan to costs of reproduction is equivocal (Hamel et al., 2010; Winder, Simons, et al., 2025). Our model predicts that the reproductive costs that are operating in early life do not cause aging directly. Aging is caused by the loss of the physiological regulation that supports Darwinian fitness returns in early life.

**Evolution of aging across species**
Different ecological conditions select for different life-histories and evolution optimizes the regulation of the underlying physiology (Stearns, 1989). The optimum in longer-lived species is shifted to retain reproduction for longer, and thus the cost curve of physiological regulation is shifted in priority. Long-lived species operate in a physiological space where they prioritize offspring quality over quantity (*K* versus *r* strategy) (Pianka, 1970). In our model we thus expect long-lived species to evolve two adaptations regarding physiological regulation. First, long-lived species will evolve more physiological regulation, to retain the optimal internal state as best as possible for as long as possible (Fedichev & Gruber, 2025), to avoid catastrophic events and produce offspring across multiple reproductive events. Second, long-lived species will operate further away from the steepest parts of the cost curve and will thus evolve more symmetrical regulation (Figure 5). Our model, in contrast to other evolutionary theories of aging, is compatible with the observation of 'immortal' species that do not show an increase in mortality risk with age (Ruby et al., 2018, 2023). In certain rare ecological conditions, it could be possible that selection in early life is so strong that it selects for such tight regulation of physiology, perhaps even resulting in a near symmetrical cost curve, that the accrued costs of physiological regulation do not result in a clear increase in mortality rate with age.

Our model leads to three predictions for the comparative biology of aging. First, longer lived organisms will be able to survive longer due to additional regulation staving off catastrophic death and frailty but will spend more time in a physiological state resembling senescence. Indeed, longer lived species spend a proportionally longer time as senescent (Ricklefs, 1998; Turbill & Ruf, 2010) and tend to show stronger fitness costs of reproductive senescence (Bouwhuis et al., 2012). Second, aging should appear more heterogeneous in long-lived species, as their cost-curve is more symmetrical, but is loaded at either end with more layers of physiological control. When these layers of regulation break during aging a larger variation in phenotypes of aging is expected. Third, pharmacological relief of costs on one side of the cost-curve is predicted to be less effective in long-lived versus short-lived species, but interventions that restore physiological regulation more generally should prove more effective in long-lived species.

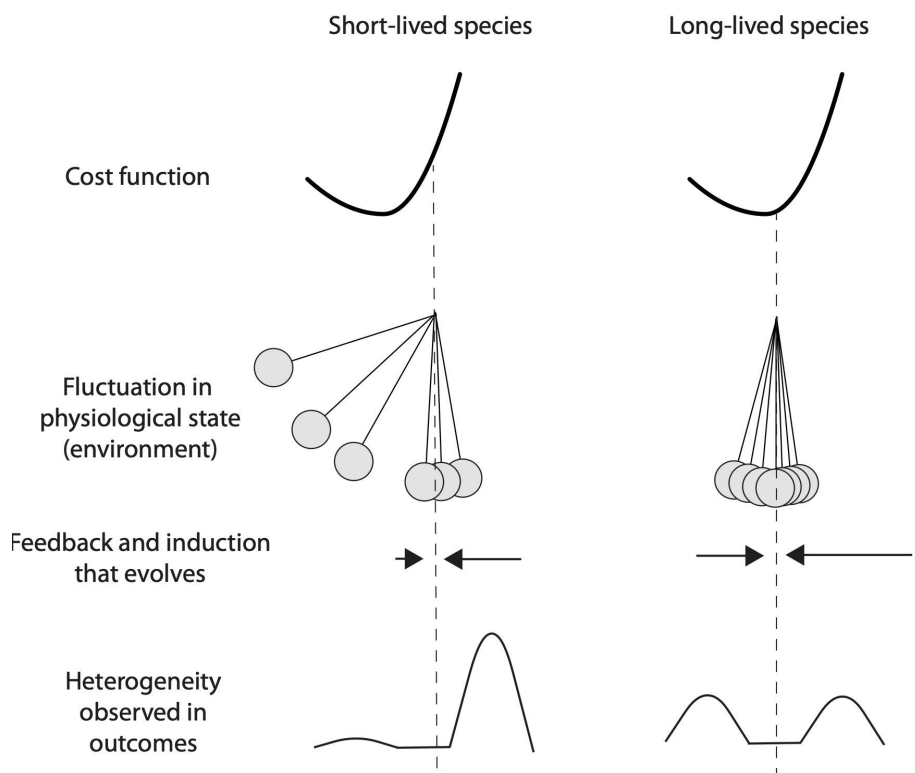

**Figure 5.** Between-species ecological niches select for different life-histories. Shorter-lived species tolerate higher costs as reproduction is favoured over direct and accrued costs. They operate at a steeper end of the physiological cost curve and evolve relatively loose regulation of physiological state. Long-lived species, avoid the high-cost space and invest in quality of offspring and longevity. In addition, they evolve strong regulatory mechanisms.

**Opposite directionality in aging-related pathology**
Disease is often difficult to distinguish from aging, when the risk of disease increases with age. The geroscience hypothesis states that aging physiology underlies all age-related diseases (Austad, 2016; Kennedy et al., 2014). Our physiological model specifies this more directly. Outcomes of aging are predicted to be multifaceted, arising from two sides of multiple physiological cost functions, whereas disease is interpreted often in a directional fashion. Disease can be a change in the effective physiological range without any change in regulation, causing disease as the organism operates at a suboptimal physiological cost point. Aging can result in a similar physiological state, resulting in aging-related disease. However, the way the organisms arrive at this state is different, as they drift into this state over time and we therefore predict aging-related disease will be accompanied by a loss of regulation. In non-aging related disease (of the same exact disease diagnosis) such loss of regulation is not necessary.

For example, the increased prevalence of type II diabetes in the elderly, can be interpreted as a disease risk resulting from the inability to produce "enough" insulin. However, the ability of beta-cells to produce insulin does not decline with age or at best is highly heterogeneous (Chia et al., 2018). Many aspects of glucose metabolism change with aging, and type II diabetes as a result from aging is thus likely to be different in its etiology than that arising from obesity, for example. Similarly, these insights can explain why seemingly opposite treatments can lead to benefits. Protein restriction, hailed as a prolongevity treatment, with strong support in model organisms (Green et al., 2022), and some support in humans, contrasts with the finding (or suggestion) that protein supplementation carries benefits in the elderly (Yoshimura et al., 2025). Androgen treatment in men carries mortality costs (Windfeld-Mathiasen et al., 2024), but testosterone therapy in elderly males is suggested to carry benefits, counteracting sarcopenia (Ottenbacher et al., 2006).

**Aging is a heterogenous devil**
Our physiological model explains why outcomes with aging are heterogenous but remain canalized. Importantly, our model suggests that some aging related outcomes are on the face of it similar to disease, but we suggest the underlying mechanisms are likely more discrete. As such we do not expect all individuals in the population to respond in the same way during aging and in response to treatments, as their responses will depend on their aging state. Within a population, heterogeneous aging due to stochastic effects results in a mosaic of organisms with different but grouped symptoms of aging. Within organisms, different cells and tissues will age at heterogeneous rates (Goeminne et al., 2025; Walker & Herndon, 2010), and thus in the same organism, one particular physiological regulation system can have aged in opposite directions.

In terms of directionality, we can expect anti-aging treatments that appear to go into a similar direction to which aging drifts to still hold functional benefits. For example, on the whole organismal level there may be reduced activation of a certain pathway, but it is the rare hyperactivation that causes mortality originating from a few cells or tissues. A suppressor of this system would yield longevity benefits. Measuring physiological change with aging on a whole organism level and expecting that to counteract this average change will yield health benefits is unlikely to be successful. In contrast, as recently suggested, studying the physiological state of animals predicted to live long (either mutant or drug treated) but measured at a young age can yield important insights into how to achieve a pro-longevity physiological state (Miller et al., 2023). Based on our model we also suggest that studying how young animals resist perturbations and how physiological feedback is organized will inform us how we can strengthen these or prevent them from breaking during aging. Treatments that both enhance positive and negative feedback in physiological systems are in our model the best geroprotectors, especially in long-lived organisms.

These predictions are relevant as major signaling pathways, such as mTOR, have mostly been interpreted and manipulated in a single direction. Overactivation of mTOR is seen as the culprit of aging that is relieved by mTOR inhibitors (e.g. rapamycin), but underactivation of mTOR during aging may also occur and can carry costs to function in certain contexts. In muscle, mTORC1 appears to be overactivated chronically during aging (Markofski et al., 2015b; Tang et al., 2019), but the inability to activate mTORC1 upon resistance exercise-training has been suggested as a reason for muscle loss (sarcopenia) with age (Fry et al., 2011). In the heart mTOR is activated during aging, but mTOR activation can be adaptive in response to stress, potentially explaining why telomerase deficient mice show reduced lifespan when treated with rapamycin (Sciarretta et al., 2022) and the differences in heart function observed between different genetic murine models of mTOR signaling (Dai et al., 2023). Hyperactivated mTOR in intestinal stem cells contributes to gut aging (He et al., 2020), and similar effects have been observed in the hematopoietic niche and muscle, but effects appear opposite in the neural stem cells (Papadopoli et al., 2019).

**Testable predictions, implications, and future directions**

1) **Costs and benefits of signaling are optimized in early life with costs outweighing the benefits when aging is observed.**
When we understand how young animals regulate their physiology in response to perturbation, we can understand how these systems will age and possibly can be rewired to sustain healthy aging for longer. We expect long-lived models (e.g. long-lived mice) to be more robust to regulation or more resistant to the negative consequences of physiological regulation at a young age already. In addition, we expect stronger regulation and more symmetrical regulation in early life to be associated with reduced aging. Importantly, the challenges imposed on the physiological system should be those that are encountered with age, otherwise our predictions will be opposite. If an organism is better able to resist physiological perturbation with the purpose to improve or secure current reproduction in early life, we expect these to be associated with faster aging, as it is the asymmetrical regulation that secures fitness in early life that later results in aging. Assessing physiological regulation *in vivo* will be the challenge to test these hypotheses.

2) **Fail safe mechanisms that secure early life fitness are expected to feature prominently in the biology of aging and the majority of these are expected to be safe targets for anti-aging interventions.** Fail-safe mechanisms are expected to have evolved to secure key aspects important in early life, sacrificing aspects of physiology in the face of catastrophic loss of early life fitness. When such mechanisms secure reproductive fitness benefits and not survival or have evolved in response to environmental challenges that can be mitigated in modern society, these fail-safe mechanisms have the potential to do excess harm during aging. Importantly, when the control of fail safes break during aging they are predicted to do significant harm to the aging organism. Interestingly, the mitigation of these failsafe mechanisms is unlikely to incur substantial side effects at old age and could therefore pose excellent targets for geroscience.

3) **Heterogeneity in aging phenotypes should decrease with increasing asymmetry of regulation.** The effects of interventions that reduce or increase signaling will depend on how asymmetrical the cost curve is. Physiology that results in largely uniform, directional, and non-heterogeneous outcomes of aging is predicted to show stronger responses to directional interventions. Examining longitudinal phenotypic data from this perspective has the potential to reveal possible novel targets for geroscience, or can aid in dosage determination.

4) **The physiological cost curve of longer-lived species is more symmetrical (i.e. aging is more apparent) and therefore long-lived species will require anti-aging treatments that enhance or maintain regulation.** Directional suppression of physiology is unlikely to be as successful in long-lived species as in short-lived species. The dose response curves of pharmacological interventions for aging are expected to have a narrower range of positive outcomes for longer-lived species as there is reduced opportunity to shift the cost curve as it is more symmetrical. Therefore, treatments that operate to improve the regulation of physiology or constrain the physiological space individuals can operate in are likely to be more successful in long-lived species. Interestingly, a recent physics-based model of aging also suggests that long-lived species would most benefit from reducing internal noise (Fedichev & Gruber, 2025).

5) **Cellular signaling or more generally physiological state will be more variable with age.** There are limited longitudinal observations of variation in physiological state in our field. We expect individuals to follow a trajectory to either side of the cost curve causing frailty, and variability in certain aspects of physiology should be more predictive of certain canalized phenotypes of aging. Assessing longitudinal aspects of physiological regulation using *in vivo* sensors, behavioural observations, or omics from excretomes provide opportunity for such studies. Such approaches will also help assess the mosaic of aging across organs or cells.

**Conclusions**

Our model explains many general features of the biology of aging. We have provided several general predictions and directions for study based on our model. We further envisage that how we view the mechanisms and symptoms of aging will benefit greatly from our model and its predictions. Individual mechanisms of aging, such as described in the hallmarks of aging (López-Otín et al., 2023) can now be viewed in a new light. We do not necessarily suggest that all aging is caused by asymmetrically regulated physiology that is selected for its early life benefits. However, the lack of evidence for classic trade-off models, such as the disposable soma theory (Douglas & Dillin, 2014; Drake & Simons, 2023; Lemaître et al., 2024; Mitchell et al., 2024; Speakman & Król, 2010; Winder, Simons, et al., 2025), suggests an alternative is appropriate. Especially considering that the strict trade-off models dictate that interventions that extend lifespan must have costs in early life (Drake & Simons, 2023) and thus interventionist strategies in geroscience will incur strong side-effects. These costs of such interventions are not necessary (Lind et al., 2021), as our model also predicts.

Our models are presented in 2-dimensional space for simplicity, but we recognise that physiology is multi-dimensional with many overlapping and redundant functions. The appreciation of complexity itself is a challenge but does explain why some interventions have such wide-reaching consequences. The shift of one physiological system to avoid the steepest costs will reverberate through connected physiology and metabolism. Understanding how our model operates across scales, from molecular interactions, to cells, to tissues, to organs to whole body function, will be informed by empirical tests of our predictions. Our perspective that integrates physiological regulation into evolutionary models of aging will guide empirical research in this still growing field.

**Funding**

Mirre Simons is a Sir Henry Dale fellow (Wellcome and Royal Society: 216405/Z/19/Z). Marc Tatar was supported during the course of this work by the National Institutes of Health, USA, through the awards: AG082801, AG059563, AG06963 and AG067323.